# Application of magnetosomes in magnetic hyperthermia

N. A. Usov[1,2,3], E. M. Gubanova[3]

[1]*National University of Science and Technology «MISiS», 119049, Moscow, Russia*
[2]*Pushkov Institute of Terrestrial Magnetism, Ionosphere and Radio Wave Propagation, Russian Academy of Sciences, IZMIRAN, 142190, Troitsk, Moscow, Russia*
[3]*National Research Nuclear University "MEPhI", 115409, Moscow, Russia*

Nanoparticles – magnetosomes, synthesized in nature by magnetotactic bacteria, are very promising for use in magnetic hyperthermia for the cancer treatment. In this work using the solution of the stochastic Landau-Lifshitz equation we calculate the specific absorbed rate (SAR) in an alternating magnetic field of assemblies of magnetosome chains depending on the particle size D, the distance between particles in a chain a, and the angle of the applied magnetic field with respect to the chain axis. The dependence of SAR on the a/D ratio is shown to have a bell-shaped form with a pronounced maximum. For a dilute oriented chain assembly with optimally chosen a/D ratio a strong magneto- dipole interaction between the chain particles leads to almost rectangular hysteresis loop, and to a large SAR of the order of 400 – 450 W/g at moderate frequencies f = 300 kHz and small magnetic field amplitudes H0 = 50 – 100 Oe. The maximum SAR value only weakly depends on the diameter of the nanoparticles and the length of the chain. However, a significant decrease in SAR occurs in a dense chain assembly due to the strong magneto-dipole interaction of nanoparticles of different chains.

**PACS** 75.20.-g; 75.50.Tt; 75.40.Mg

**Keywords:** Magnetotactic bacteria; Magnetosome chain; Magnetic hyperthermia; Low frequency hysteresis loops; Numerical simulation

## Introduction

Magnetic hyperthermia [1-3] is currently considered as a very promising method of cancer treatment. Laboratory tests [4-7] and clinical studies [8-10] show that local, dosed heating of tumors leads to a delay in their growth and even complete decay and disappearance. For this, it is necessary in several sessions to maintain the temperature of the tumor at the level of 43-45ºC for 30 minutes [1-3]. In magnetic hyperthermia local heating of biological tissues is achieved by introducing magnetic nanoparticles into a tumor and applying to them a low frequency alternating magnetic field [1-3,11,12]. The fundamental advantage of magnetic hyperthermia lies in the possibility of local heating of tumors located deep in a human body, since the low-frequency magnetic field is relatively poorly shielded by the biological medium, in contrast to high-frequency and laser heating methods [13,14], which are suitable for the treatment of tumors localized near body surface.

However, for the successful implementation of the magnetic hyperthermia in practice it is necessary to use assemblies of magnetic nanoparticles with a sufficiently high specific absorption rate (SAR) in alternating magnetic field of moderate frequency, $f$ <1 MHz, and amplitude $H_0$ of the order of 100 - 200 Oe. It is proved [15,16] that the effect of an alternating magnetic field on living organisms is relatively safe under the condition $H_0 f \leq 5 \times 10^9$ A(ms)$^{-1}$. In addition, only mildly toxic and biodegradable nanoparticles can be used in magnetic hyperthermia to reduce the risk of possible side effects. Based on these requirements, magnetic nanoparticles of iron oxides are considered to be the most suitable for application in magnetic hyperthermia [1-3,11,12,16-19].

Unfortunately, the popular chemical methods for the synthesis of magnetic nanoparticles of iron oxides [1, 20–22] in most cases give assemblies with a wide distribution of nanoparticles in size and shape. Moreover, the obtained nanoparticles turn out to be polycrystalline, which leads to a low saturation magnetization of particles in comparison with the corresponding single-crystal three-dimensional samples [23]. For such assemblies it is hardly possible to obtain sufficiently high SARs under the above restrictions on the frequency and amplitude of alternating magnetic field [24.25]. Furthermore, the subsequent uncontrolled agglomeration of nanoparticles in the biological medium leads to a further decrease of the assembly SAR [26,27].

The results of numerical simulations [28–30], confirmed by a number of experimental data [25, 31–36] show that in order to achieve sufficiently high SARs at moderate amplitudes and frequencies of alternating magnetic field, one has to use single-crystal magnetic nanoparticles with high saturation magnetization. It is necessary also to ensure a narrow particle size distribution near the optimally chosen nanoparticle diameter. Due to difficulties with chemical synthesis, lately much attention has been paid to experimental and theoretical studies of assemblies of magnetosomes, which are synthesized in nature by magnetotactic bacteria [5,25,37-44]. Magnetosomes grow inside bacteria under optimal physiological conditions. Therefore, they have a perfect crystalline structure, a quasi-spherical shape, and a fairly narrow particle size distribution. Most magnetotactic bacteria synthesize nanoparticles that are close in chemical composition to high-purity maghemite, γ-$Fe_2O_3$ [41] or magnetite, $Fe_3O_4$, [44].

It was shown in the pioneer work of Hergt, *et. al.* [37] that the SAR in the oriented assembly of magnetosomes reaches a very high value of 960 W/g at a frequency $f$ = 410 kHz and magnetic field amplitude $H_0$ = 126 Oe. High SAR values in magnetosome assemblies were also obtained in a number of subsequent experimental studies [5, 25, 44]. It is known [37–39, 42] that various types of magnetotactic bacteria synthesize nanoparticles with different characteristic sizes, from 20 to 50 nm. In the bacteria the nanoparticles are arranged in the form of long chains consisting of 6 to 30 particles of approximately the same diameter. Existing experimental techniques make it possible to isolate magnetosomes from bacteria both in the form of single particles and in the form of chains with different numbers of nanoparticles in the chain [25, 37–42].

The behavior of the assembly of magnetosome chains in an alternating magnetic field is of great interest from a theoretical point of view. It was already emphasized earlier [45–47] that the orientation of individual nanoparticles and their chains along the applied alternating magnetic field is important to obtain appreciable SARs for magnetosome assembly. In this paper, based on the solution of the stochastic Landau – Lifshitz equation [48-51], the low-frequency hysteresis loops of assemblies of magnetosome chains extracted from magnetotactic bacteria are calculated. This theoretical approach makes it possible to take into account the complicated magnetic anisotropy of particles, the influence of thermal fluctuations of magnetic moments, and strong magnetic dipole interaction between chain particles. In this work the dependence of the assembly SAR on the particle diameter, on the length of the chains, and the distance between particles in the chain is studied in detail. We investigate also the dependence of SAR on the orientation of applied magnetic field with respect to the chain axis, as well as on the average density of the chain assembly. It is shown that with an optimal choice of magnetosome chain geometry, sufficiently high SAR values, of the order of 400 - 450 W/g, can be obtained at a frequency $f$ = 300 kHz and at small and moderate field amplitudes, $H_0$ = 50 - 100 Oe. However, it is found that the SAR of the chain assembly decreases significantly with an increase in its average density. It seems that the results obtained will be useful for the optimal choice of the geometric parameters of magnetosome chains to further increase their heating ability for successful application in magnetic hyperthermia.

**Numerical Simulation**

Calculations of SAR of magnetosome chains were performed in this work by numerical simulation using stochastic Landau- Lifshitz equation [48-51]. The latter governs the dynamics of the unit magnetization vector $\vec{\alpha}_i$ of *i*-th single-domain nanoparticle of the magnetosome chain

$$\frac{\partial \vec{\alpha}_i}{\partial t} = -\gamma_1 \vec{\alpha}_i \times (\vec{H}_{ef,i} + \vec{H}_{th,i}) - \kappa \gamma_1 \vec{\alpha}_i \times (\vec{\alpha}_i \times (\vec{H}_{ef,i} + \vec{H}_{th,i})), \quad i = 1,2,..N_p. \quad (1)$$

Here $\gamma$ is the gyromagnetic ratio, $\kappa$ is phenomenological damping parameter, $\gamma_1 = \gamma/(1+\kappa^2)$, $\vec{H}_{ef,i}$ is the effective magnetic field, $\vec{H}_{th,i}$ is the thermal field, and $N_p$ is the number of nanoparticles in the chain. The effective magnetic field acting on a separate nanoparticle can be calculated as a derivative of the total chain energy

$$\vec{H}_{ef,i} = -\frac{\partial W}{M_s V \partial \vec{\alpha}_i}. \quad (2)$$

The total magnetic energy of the chain $W = W_a + W_Z + W_m$ is a sum of the magnetic anisotropy energy $W_a$, Zeeman energy $W_Z$ of the particles in applied magnetic field, and the energy of mutual magneto-dipole interaction of the particles $W_m$. Since magnetosomes released from bacteria are coated with thin non-magnetic shells, there is no direct contact between the magnetic cores of the nanoparticles. Therefore, the exchange interaction of neighboring nanoparticles in the chain can be neglected.

The magneto- crystalline anisotropy energy of magnetosomes with perfect crystalline structure is that of cubic type

$$W_a = K_c V \sum_{i=1}^{N_p} \left( (\vec{\alpha}_i \vec{e}_{1i})^2 (\vec{\alpha}_i \vec{e}_{2i})^2 + (\vec{\alpha}_i \vec{e}_{1i})^2 (\vec{\alpha}_i \vec{e}_{3i})^2 + (\vec{\alpha}_i \vec{e}_{2i})^2 (\vec{\alpha}_i \vec{e}_{3i})^2 \right)$$
(3)

where ($\vec{e}_{1i}$, $\vec{e}_{2i}$, $\vec{e}_{3i}$) is a set of orthogonal unit vectors that determine an orientation of *i*-th nanoparticle of the chain. It is assumed that the easy anisotropy axes of various nanoparticles in the chain are randomly oriented with respect to each other.

Zeeman energy of the chain in applied alternating magnetic field $\vec{H}_0 \sin(\omega t)$ is given by

$$W_Z = -M_s V \sum_{i=1}^{N_p} (\vec{\alpha}_i \vec{H}_0 \sin(\omega t)). \quad (4)$$

For nearly spherical uniformly magnetized nanoparticles the magnetostatic energy of the chain can be represented as the energy of the point interacting dipoles located at the particle centers $\vec{r}_i$ within the chain. Then the energy of magneto-dipole interaction is

$$W_m = \frac{M_s^2 V^2}{2} \sum_{i \neq j} \frac{\vec{\alpha}_i \vec{\alpha}_j - 3(\vec{\alpha}_i \vec{n}_{ij})(\vec{\alpha}_j \vec{n}_{ij})}{|\vec{r}_i - \vec{r}_j|^3}, \quad (5)$$

where $\vec{n}_{ij}$ is the unit vector along the line connecting the centers of *i*-th and *j*-th particles, respectively.

The thermal fields $\vec{H}_{th,i}$ acting on various nanoparticles of the chain are statistically independent, with the following statistical properties [48] of their components

$$\langle H_{th}^{(\alpha)}(t) \rangle = 0;$$
$$\langle H_{th}^{(\alpha)}(t) H_{th}^{(\beta)}(t_1) \rangle = \frac{2k_B T \kappa}{\gamma M_s V} \delta_{\alpha\beta} \delta(t - t_1),$$
$$\alpha, \beta = (x, y, z). \qquad (6)$$

Here $k_B$ is the Boltzmann constant, $\delta_{\alpha\beta}$ is the Kroneker symbol, and $\delta(t)$ is the delta function. The numerical simulation procedure is described in details in Refs. 58, 59.

## Results

In this work the calculations of SAR of magnetosome chain assemblies are performed in the frequency range $f$ = 250 - 350 kHz, at small and moderate amplitudes of an alternating magnetic field, $H_0$ = 50 - 150 Oe, since the use of alternating magnetic fields of small amplitude is preferable in a medical clinic.

### 3.1. Dilute chain assembly

Let us consider first the properties of dilute oriented assemblies of magnetosome chains neglecting the magnetic dipole interaction between various chains. It is of considerable interest to study the dependence of the SAR of such an assembly on the nanoparticle diameter $D$, on the number of particles in the chain $N_p$, and on the orientation of the external magnetic field with respect to the chain axis. For completeness, the chains with different average distances $a$ between the centers of the nanoparticles are considered. Note that all geometric parameters mentioned can be adopted properly during the experimental design of a chain from individual magnetosomes of approximately the same diameter [5,7,39-41]. In particular, the distance $a$ between the centers of successive nanoparticles in a chain is determined by the thickness of non-magnetic shells on their surfaces. The latter protect the nanoparticles from the aggressive action of the medium. It is worth noting that since the cubic magnetic anisotropy constant of magnetite is negative, $K_c$ = -10$^5$ erg/cm$^3$, the quasi-spherical magnetite nanoparticles have eight equivalent directions of easy anisotropy axes [52]. In the calculations performed it is assumed that the easy anisotropy axes of individual magnetosomes are randomly oriented, since it is hardly possible to make multiple easy anisotropy axes of various magnetosomes parallel when a chain is created. Saturation magnetization of the magnetosomes is assumed to be $M_s$ = 450 emu/cm$^3$ [25,44]. It is supposed also that magnetosome chain can not rotate as a whole being distributed in a medium with a sufficiently high viscosity.

Figure 1a shows the results of SAR calculation for dilute oriented assemblies of magnetosome chains with particles of various diameters, $D$ = 20 - 50 nm. The number of particles in the chains is fixed at $N_p$ = 30, alternating magnetic field frequency $f$ = 300 kHz, field amplitude $H_0$ = 50 Oe. The alternating magnetic field is applied parallel to the chain axis, the easy anisotropy axes of various particles in the chain are randomly oriented. The calculation results are averaged over a sufficiently large number of independent chain realizations, $N_{exp}$ = 40 – 60. The temperature of the system is $T$ = 300°K.

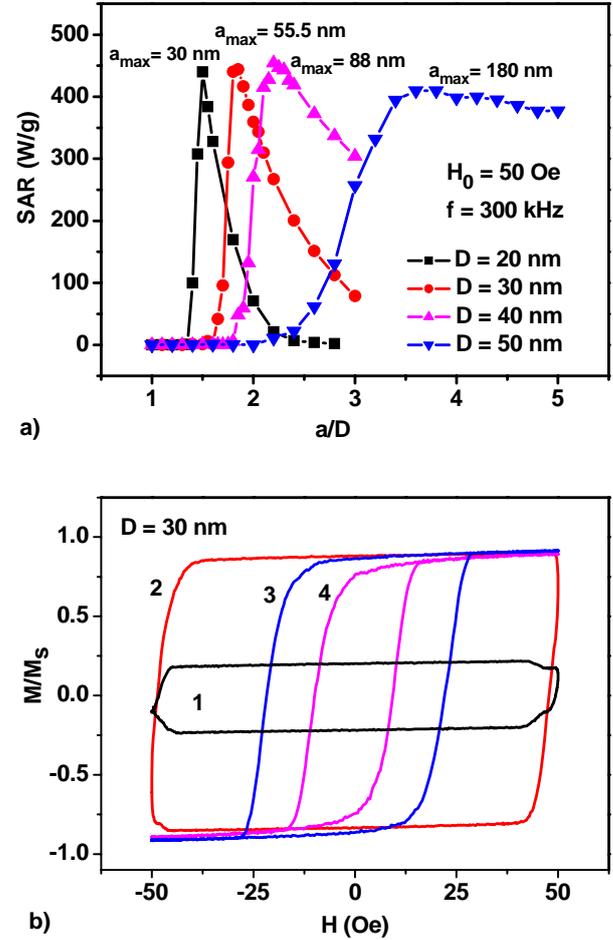

**Figure 1**. a) Dependence of assembly SAR on the reduced distance $a/D$ between the particle centers for chains with various particle diameters. b) Evolution of the shape of the low-frequency hysteresis loop for chains of particles with diameter $D$ = 30 nm for various reduced distances: 1) $a/D$ = 1.7, 2) $a/D$ = 1.85, 3) $a/D$ = 2.4, 4) $a/D$ = 3.0.

As Figure 1a shows, in all cases considered there is a significant dependence of SAR on the reduced distance $a/D$ between the centers of the particles of the chain. The assembly SAR reaches a maximum at the ratios $a/D$ = 1.5, 1.85, 2.2, and 3.6 for particles with diameters $D$ = 20, 30, 40, and 50 nm, respectively. The SAR decreases sharply after reaching the maximum. It is interesting to note that the SAR at the maximum only weakly depends on the particle diameter. Indeed, according to Figure 1a for particles with diameters $D$ = 20, 30, 40, and 50 nm, the maximum SAR values are given by 440, 444, 454, and 409 W/g, respectively.

Figure 1b shows the evolution of the low frequency hysteresis loops of a dilute chain assembly with nanoparticle diameter $D$ = 30 nm as a function of the reduced distance $a/D$ between the nanoparticle centers. It is known [28, 29] that the SAR of an assembly of magnetic nanoparticles is proportional to the area of the low frequency hysteresis loop. From Figure 1b it can be seen that in accordance with Figure 1a the maximum area of the hysteresis loop for particles with diameter $D = 30$ nm corresponds to the reduced distance $a/D = 1.85$.

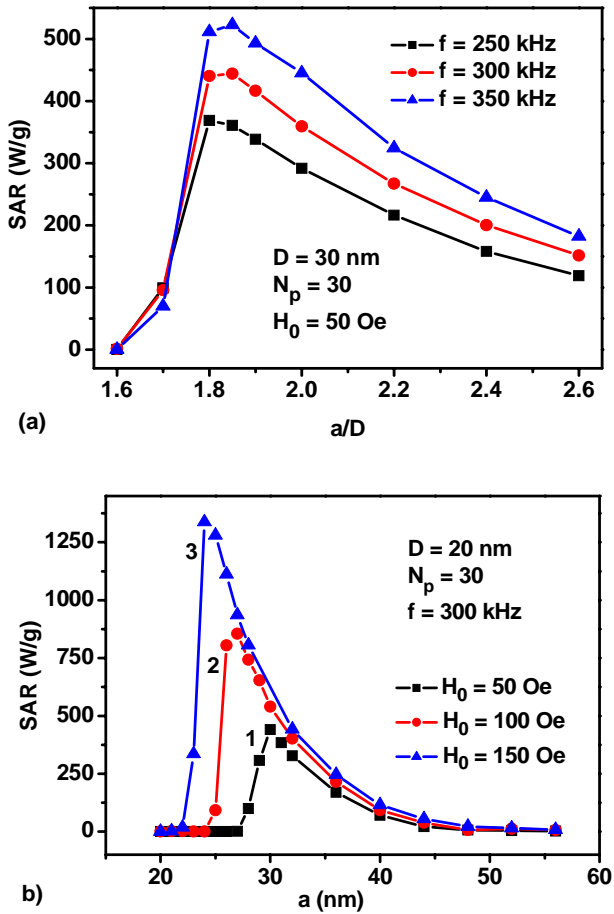

**Figure 2**. a) Frequency dependence of SAR for assembly of magnetosome chains with particle diameter $D = 30$ nm in an alternating magnetic field with an amplitude $H_0 = 50$ Oe. b) Dependence of SAR on the distance $a$ between the particle centers for particles with diameter $D = 20$ nm for various amplitudes of alternating magnetic field: 1) $H_0 = 50$ Oe, 2) $H_0 = 100$ Oe, 3) $H_0 = 150$ Oe.

It is interesting to note that the position of the SAR maximum is practically independent on the frequency. For example, according to Figure 2a for the case of particles with diameter $D = 30$ nm the SAR maximum corresponds to the ratio $a/D \approx 1.8$ in the frequency range $f = 250 - 350$ kHz. At the same time, as Figure 2b shows, the position of the maximum and the SAR value at the maximum substantially depend on the amplitude of the alternating magnetic field. Therefore, the choice of the optimal $a/D$ ratio for a magnetosome chain should be consistent with the given value of $H_0$. With increasing $H_0$ the maximum of the assembly SAR grows, and the position of the maximum falls at a shorter distance between the particles of the chain. Indeed, in Figure 2b the maximum values of SAR = 440.1, 854.6, and 1281.0 W/g are observed at $a/D = 1.5$, 1.35, and 1.2 for the magnetic field amplitudes $H_0 = 50$, 100, and 150 Oe, respectively. In general, as Figures 1 and 2 show, for a dilute oriented assembly of magnetosome chains with an optimal choice of the $a/D$ ratio one can obtain rather high SAR values, of the order of 400 - 450 W/g already in an alternating magnetic field with a relatively small amplitude $H_0 = 50$ Oe.

The significant dependence of the assembly SAR on the $a/D$ ratio, and on the amplitude of the ac magnetic field $H_0$, shown in Figures 1a and 2b, is explained by the influence of a strong interacting field $H_d$ acting between closely located particles of the chain. Figure 3a shows the instantaneous distribution of interacting fields for an individual chain consisting of 30 nanoparticles of diameter $D = 20$ nm at a time when the alternating magnetic field is close to zero. The $Z$ axis is assumed to be parallel to the axis of the chain, the amplitude and frequency of the alternating magnetic field are $H_0 = 50$ Oe, and $f = 300$ kHz, respectively. As Figure 3a shows, in the central part of the chain, due to the summation of the magnetic fields of individual nanoparticles, the longitudinal component of the interacting field reaches sufficiently large values, $H_{dz} = 250$ Oe, significantly exceeding the amplitude of the alternating magnetic field, $H_0 = 50$ Oe. On the contrary, the transverse field components turn out to be relatively small $|H_{dx}|$, $|H_{dy}|$ < 50 Oe. Obviously, the irregular distribution of the interacting field on individual particles, shown in Figure 3a is associated with a random orientation of the easy anisotropy axes of individual nanoparticles.

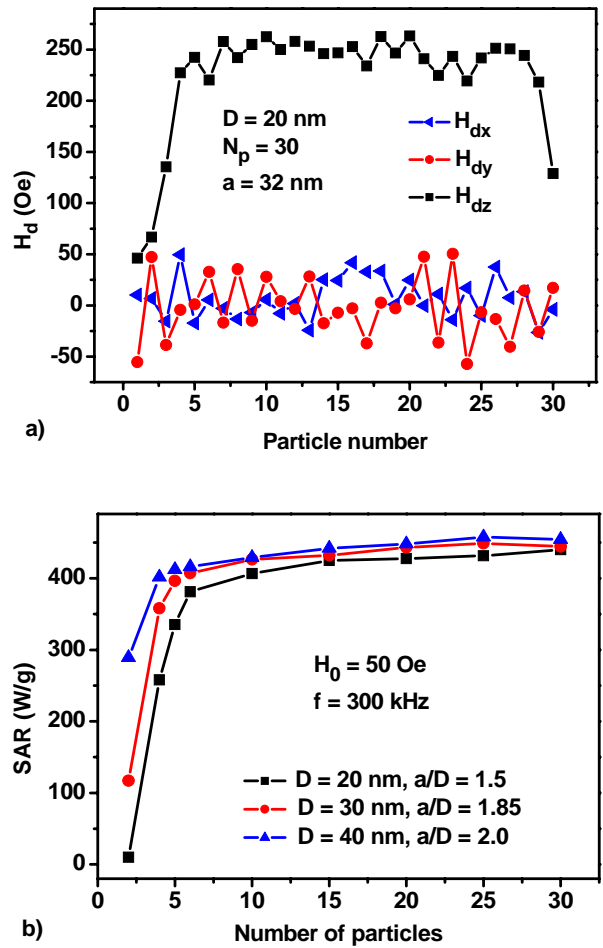

**Figure 3**. a) Distribution of the components of the interaction field depending on the position of the nanoparticle in the chain. b) Dependence of the SAR of a dilute chain assembly on the number of particles $N_p$ in the chains with particles of different diameters, $D = 20–40$ nm, assuming the optimal ratios $a/D$ for various chains.

As Figure 3a shows the interacting field is greatly reduced at the chain ends. Therefore, the magnetization switching field $H_{sf}$ of the chain is determined by the conditions near its ends. Calculations show that for $H_{sf} < H_0$ the chain magnetization reverses as a whole in a process similar to the giant Barkhausen jump in iron-rich amorphous ferromagnetic microwires [53]. The switching field $H_{sf}$ of the chain increases with decreasing average distance between the particles in the chain, since the intensity of the magnetic dipole interaction increases if neighboring nanoparticles are located closer to each other. When the distance between the particle centers decreases, the switching field of the chain can reach values exceeding the amplitude of alternating magnetic field, $H_{sf} > H_0$. Under this condition the magnetization reversal of the chain is impossible. As Figure 1b shows, when the reduced distance between the particle centers decreases from $a/D = 1.85$ to $a/D = 1.7$, the area of the assembly hysteresis loop reduces sharply. This leads to a sharp drop in the SAR of the corresponding assembly in Figure 1a. However, the drop in the hysteresis loop area in Figure 1b does not occur immediately to zero, due to fluctuations in the switching fields $H_{sf}$ of individual chains of the assembly. As a result, even at $a/D = 1.7$ some of the assembly chains are still able to reverse their magnetizations in applied magnetic field. This, however, leads to a sharp narrowing of the vertical size of the low frequency hysteresis loop of the whole assembly. With further decrease of the $a/D$ ratio the fraction of chains capable of magnetization reversal at given magnetic field amplitude $H_0$ tends to zero. As a result, the ability of the assembly to absorb the energy of an alternating magnetic field disappears.

Similar considerations explain the behavior of the assembly SAR as a function of the alternating magnetic field amplitude. As Figure 2b shows, with an increase in $H_0$ the magnetization reversal of the chain is possible at smaller reduced distances $a/D$. This leads to a shift in the positions of the SAR maximum to smaller reduced distances and to increase in the maximum SAR values.

It is important to note the characteristic change of the shape of the low frequency hysteresis loop of a dilute magnetosome assembly as a function of $a/D$ ratio. As Figure 1b shows, the hysteresis loop is nearly rectangular and has maximal area for optimal ratio $(a/D)_0$. For ratios smaller than optimal, $a/D < (a/D)_0$, the vertical size of the hysteresis loop is considerably decreased, $M/M_s < 1$. On the other hand, for ratios larger than optimal, $a/D > (a/D)_0$, the width of the assembly hysteresis loop is reduced, $2H_{max} < 2H_0$.

For an infinite periodic chain of single-domain nanoparticles with uniaxial magnetic anisotropy, the switching field for magnetization reversal was analytically determined by Aharoni [54]. However, in the present case an analytical consideration is hardly possible due to the randomness in the orientation of the easy anisotropy axes of individual nanoparticles and the influence of thermal fluctuations of the particle magnetic moments at a finite temperature.

It is interesting to note that the effect of sharp variation of the SAR of a dilute assembly of magnetosome chains as a function of the $a/D$ ratio only slightly depends on the number of particles in the chains, provided that it exceeds the value of the order of $N_p = 4$-6. For example, Figure 3b shows the dependences of SAR on the number of particles, $N_p = 2$–30, in the chain assemblies with particles of various diameters, $D = 20 – 40$ nm. The calculations presented in Figure 3b are performed at optimal ratios $a/D = 1.5$, 1.85, and 2.2 for particles with diameters $D = 20$, 30, and 40 nm, respectively. These optimal $a/D$ values were determined previously at frequency $f = 300$ kHz and amplitude $H_0 = 50$ Oe. According to Figure 3b, regardless of the nanoparticle diameter, a sharp increase in SAR as a function of the number of particles in the chains occurs in the range $N_p \leq 4 - 5$. However, with further increase in the chain length the SARs of the assemblies change slowly.

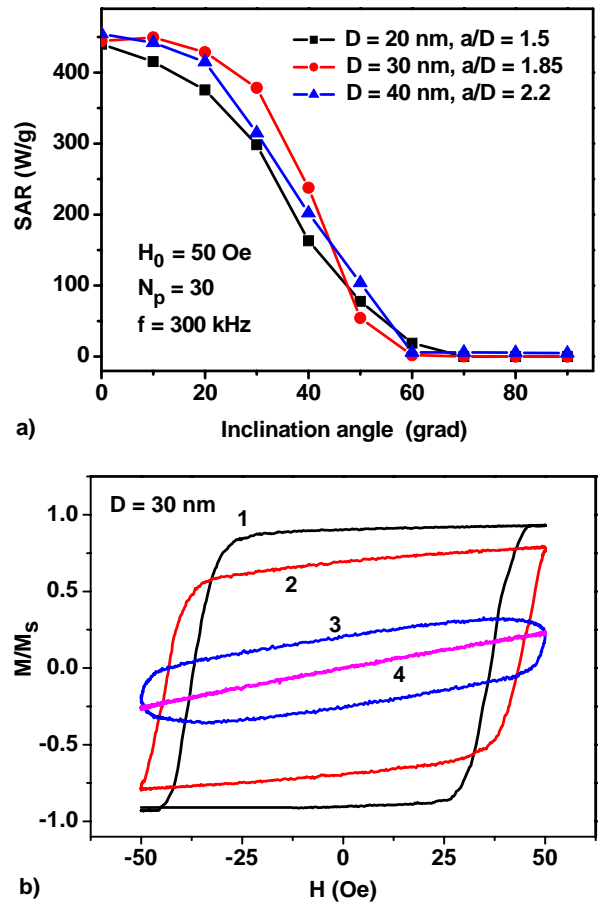

**Figure 4.** a) Dependence of the assembly SAR on the angle of alternating magnetic field with respect to the chain axis for assemblies of particles of different diameters. b) The evolution of low frequency hysteresis loops for assembly of particles with diameter $D = 30$ nm for different directions of applied magnetic field with respect to the common chain axis: 1) $\theta = 0$, 2) $\theta = 40º$, 3) $\theta = 60º$, 4) $\theta = 80º$.

Let us consider now the dependence of the SAR of a dilute oriented chain assembly on the angle $\theta$ of alternating magnetic field with respect to the chain axis. The calculations presented in Figure 4 are carried out at frequency $f = 300$ kHz and magnetic field amplitude $H_0 = 50$ Oe, the number of particles in the chains being $N_p = 30$. The reduced distances between the centers of neighboring nanoparticles in the chains are chosen optimal, so that $a/D = 1.5, 1.85$, and $2.2$ for particles with diameters $D = 20, 30$, and 40 nm, respectively.

As Figure 4a shows, the maximum SAR value is achieved when the alternating magnetic field is parallel to the chain axis, $\theta = 0$. It is remarkable, however, that in a rather wide range of angles, $\theta \leq 30 - 40º$, the SAR of the assembly varies slightly. Only for angles $\theta > 50º$ the assembly SAR drops sharply, so that for magnetic field directed perpendicular to the chain axis the assembly SAR is close to zero. Figure 4b shows the evolution of the low frequency hysteresis loops for an assembly of nanoparticles with a diameter of $D = 30$ nm as a function of $\theta$. In accordance with Figure 4a, the hysteresis loop of the maximum area corresponds to the angle $\theta = 0$, and for angles $\theta \geq 60º$ the hysteresis loop area tends to zero.

The fact that the angular dependence of assembly SAR turns out to be rather slow in the range of angles $\theta \leq 30 - 40º$ means that the results presented in Figures 1-3 are approximately true for dilute partially oriented assemblies of magnetosome chains. Our calculations also show that a small variation in particle diameters within the same chain, random variations in the distances between the centers of particles in the chain, and small random deviations of the particle positions from the straight line do not significantly affect the low frequency hysteresis loops averaged over a representative assembly of magnetosome chains.

### 3.2. Interaction of magnetosome chains

In this section the results obtained above are generalized to the case of dilute assembly of clusters of magnetosome chains of various densities. Inside the cluster due to the close arrangement of the chains the magnetic dipole interaction of nanoparticles of various chains should be taken into account. Figure 5a shows the results of calculating the low frequency hysteresis loops of an assembly of oriented cylindrical clusters with an overall diameter $D_{cl} = 280$ nm, and a height $L_{cl} = 480$ nm. The axis of the chains is parallel to the axis of the cylinder, but the positions of the chains in the cluster are distributed randomly, as shown schematically in Figure 5b. The alternating magnetic field is applied along the chain axis. In the calculations performed it is assumed that the centers of the nanoparticles in the chains are located at an optimal distance, $a = 2.2D$, for nanoparticles with diameter $D = 40$ nm. This allows one to compare the numerical results with the data shown in Figure 1a for a dilute assembly of individual chains with the same nanoparticle diameter. The frequency and amplitude of the alternating magnetic field are $f = 300$ kHz and $H_0 = 50$ Oe, respectively.

In the illustrative calculations performed, the number of particles in the chains is fixed at $N_p = 6$, but the number of chains $N_{ch}$ in a cluster of given diameter $D_{cl} = 280$ nm varied from 4 to 10. In this manner we can change the total number of nanoparticles in the cluster, $N_p N_{ch}$, as well as the cluster filling density, $\eta = N_p N_{ch} V/V_{cl}$, where $V = \pi D^3/6$ is the volume of the nanoparticle, $V_{cl} = \pi D_{cl}^2 L_{cl}/4$ being the volume of the cylindrical cluster. SAR calculations of dilute assemblies of clusters with different numbers of chains $N_{ch}$ are averaged over a sufficiently large number of independent realizations of random clusters, $N_{exp} = 40 - 60$.

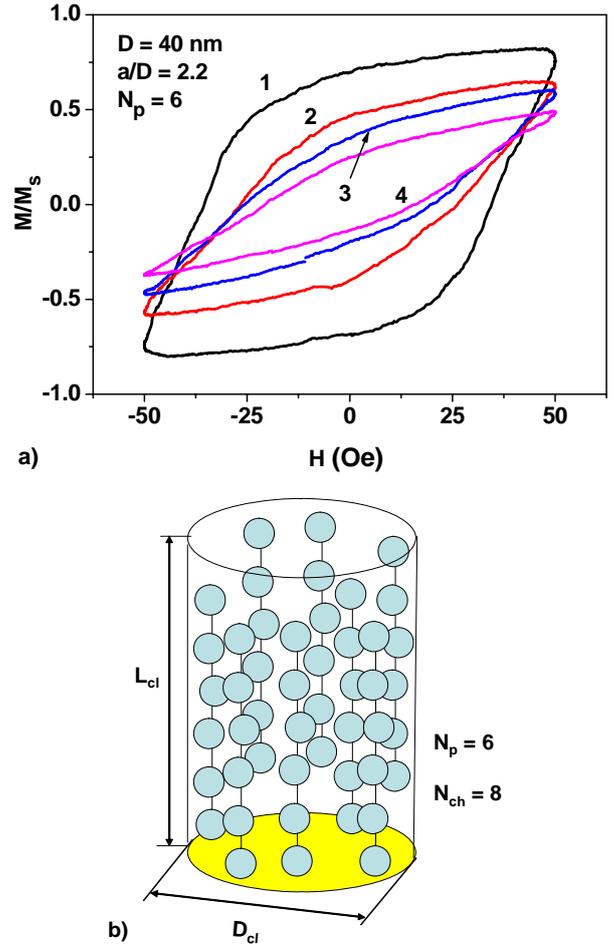

**Figure 5**. a) Low-frequency hysteresis loops of dilute assemblies of cylindrical clusters of interacting chains of nanoparticles depending on the number of chains located inside the cluster: 1) $N_{ch} = 4$, 2) $N_{ch} = 6$, 3) $N_{ch} = 8$, 4) $N_{ch} = 10$. b) The model of random oriented cluster of magnetosome chains used in the calculations.

As Figure 5a shows, due to an increase in the intensity of the magnetic dipole interaction inside the clusters, the area of the assembly hysteresis loop rapidly decreases with an increase in the cluster filling density. The cluster filling density increases as $\eta = 0.027, 0.041, 0.054$, and $0.068$ for the clusters of the given size with the number of chains within the cluster $N_{ch} = 4, 6, 8$, and 10, respectively. It is found that with increase in the cluster filling density, the SAR of the assembly of clusters rapidly decreases as follows: SAR = 270.9, 145.5, 98.2, and 62.5 W/g. It should be noted that for a dilute assembly of non-interacting chains (see Figure 1a) with the same chain geometry, frequency and amplitude of the alternating magnetic field, the SAR of the assembly at maximum reaches the value 454 W/g for the particles with optimal ratio $a/D$ = 2.2. Thus, a significant drop in the SAR assembly due to the magnetic dipole interaction of individual chains must be taken into account when analyzing experimental data.

**Discussion**

As emphasized above, the properties of magnetosome assemblies are currently being actively studied for application in biomedicine [7,25,37–47]. In particular, SARs of magnetosome assemblies are measured [25,37,40,41,44] in an alternating magnetic field under various conditions. In addition, assemblies of magnetosome chains are used in laboratory experiments [7,38,39,41] for the successful treatment of glioblastoma in mice using magnetic hyperthermia.

Low-frequency hysteresis loops of dilute assemblies of isolated magnetosomes with an average diameter $D$ = 45 ± 6 nm were measured [25] at selected frequencies of 75 kHz, 149 kHz, 302 kHz and 532 kHz, in the range of magnetic fields up to 570 Oe. The measurements were carried out for magnetosome assemblies distributed in water, and in agarose gel with an increased viscosity. It was shown that in both cases the SAR of the assemblies almost linearly depend on the alternating field frequency. For the assembly of magnetosomes dispersed in water, a very high value of SAR = 880 W/g was obtained at a frequency $f$ = 200 kHz and magnetic field amplitude $H_0$ = 310 Oe. In agarose gel the SAR of the assembly at the same frequency and magnetic field amplitude turns out to be somewhat lower, SAR = 635 W/g, since in a medium with increased viscosity Brownian contribution to SAR is significantly reduced. These data are in agreement with the high SAR values obtained previously for magnetosome assemblies by Hergt *et. al.* [37].

Recently [44] very large SARs were obtained by the same authors for assemblies of "whole" magnetotactic bacteria distributed in water. The importance of the orientation of the chains of magnetosomes located inside the bacteria in the direction of alternating magnetic field has been emphasized again. Since the chains of magnetosomes are located inside bacteria with sizes 2 – 5 by 0.5 μm, they turn out to be efficiently separated by organic material. This reduces the intensity of the magneto-dipole interaction of nanoparticles belonging to different chains. It is shown [44] that the heating efficiency of oriented assembly of magnetotactic bacteria is appreciably higher than the one obtained [25] for the assembly of isolated magnetosomes. Actually, the SAR of the assembly of magnetotactic bacteria measured by means of AC magnetometry reaches the huge value of 2400 W/g at frequency $f$ = 300 kHz and magnetic field amplitude $H_0$ = 380 Oe. Nevertheless, it seems hardly possible to uniformly distribute and correctly orient within the tumor large bacteria of 2-5 μm in length. This can probably be done much easier using short chains of magnetosomes having 4 - 6 nanoparticles in length. For example, the optimal length of a chain consisting of 5 nanoparticles with diameter $D$ = 20 nm is given by only 140 nm. In addition, when working with chains created from individual magnetosomes, it becomes possible to optimize the geometry of the chains, that is, to ensure the optimal $a/D$ ratio, matching it with the applied value of $H_0$.

In this regard, a technique for working with magnetosomes developed by E. Alphandery with co-workers [7,38–41] seems to be promising. They removed most of the organic materials, including endotoxins, from magnetosomes extracted from magnetotactic bacteria. The nanoparticles are then stabilized with different bio-degradable and biocompatible coating agents of various thickness, such as poly-L-lysine, oleic acid, citric acid, or carboxy-methyl-dextran [40]. Then, individual magnetosomes coated with these shells are assembled into chains with a small number of particles, $N_p$ = 4 - 6, and introduced into the tumor for magnetic hyperthermia. A typical concentration of magnetic nanoparticles introduced into a tumor is given by 25 μg in iron of nanoparticles per mm$^3$ of tumor [7,40,41]. Taking into account the density of magnetosomes around 5 g/cm$^3$, and assuming nearly uniform distribution of the magnetosome chains within the tumor, one obtains the average density of the assembly $\eta$ = 0.005. Such chain assembly can be considered quite diluted, so that the results obtained in Section 3.1 above can be applied for the theoretical estimation of the assembly SAR.

However, Le Fèvre *et. al.* [41] obtained rather small SAR values for the assembly of the magnetosome chains studied, only 40 W/g$_{Fe}$ at frequency $f$ = 198 kHz. As a result, in order to obtain the required tumor temperature of 43-46° C, it was necessary to use rather large amplitudes of the alternating magnetic field, $H_0$ = 110 - 310 Oe. Similarly, only relatively small values of SAR = 89 - 196 W/g$_{Fe}$ were obtained in Ref. 40 at frequency $f$ = 198 kHz at sufficiently large magnetic field amplitudes, $H_0$ = 340 - 470 Oe. This may indicate that the geometric structure of the magnetosome chains, in particular, the $a/D$ ratio, was not optimal in these experiments. Indeed, the individual magnetosomes used [41] to construct the magnetosome chains were covered by rather thin poly-L-lysine shells with a thickness of $t$ = 4 - 17 nm. Therefore, for a significant fraction of the magnetosome chains with an average particle diameter $D$ = 40.5 ± 8.5 nm, the ratio $a/D$ = 1 + 2$t/D$ ≈ 1.2 – 1.6 may turn out to be far from the optimal value for the amplitudes of the alternating magnetic field used. Similarly, magnetosomes with an average diameter $D$ = 40–50 nm were covered with shells of various chemical compositions [40], but of sufficiently small thickness, $t$ = 2 – 6 nm.

Another reason for the small SAR values observed in experiments [40, 41] may be associated with the formation of dense clusters of magnetosome chains inside the tumor. As shown in Section 3.2, in dense cluster of magnetosome chains a significant decrease in SAR occurs due to the strong magneto-dipole interaction of nanoparticles of different chains. Finally, as shown in Section 3.1, the unfavorable orientation of the chains with respect to the direction of applied magnetic field can also cause a significant drop in the assembly SAR. In this regard, it would be interesting to check the possible effect of a strong magnetic field of a Neodymium magnet on the orientation of short magnetosome chains ($N_p$ = 4-6) within a tumor. A constant magnetic field can be temporarily applied during the procedure of introducing magnetosome chains into a tumor to provide correct chain orientation.

## Conclusions

It is now widely accepted [1-3] that for use in magnetic hyperthermia it is extremely important to ensure a high quality of magnetic nanoparticles. For dilute assemblies of single-domain magnetic nanoparticles with perfect crystalline structure, sufficiently high saturation magnetization and narrow particle size distribution, sufficiently high SAR can be obtained [25, 31–35] at moderate frequencies and amplitudes of an alternating magnetic field satisfying Brezovich [15] or Hergt *et. al.* criteria [16]. These experimental results are in satisfactory agreement with the earlier theoretical estimates [28,29]. However, in many experimental studies it was found [26,27,55-57] that when an assembly of magnetic nanoparticles is introduced into a tumor, uncontrolled agglomeration of nanoparticles occurs with the formation of dense clusters of nanoparticles of various geometric structures. As a result, the SAR of the assembly significantly decreases [26,27,55] due to the influence of a strong magnetic dipole interaction between the nanoparticles of the cluster. Recent theoretical calculations show [58-59] that with an increase in the average density of a nanoparticle assembly in the range $\eta$ = 0.01–0.35, the maximum SAR of the assembly of nanoparticles with different types of magnetic anisotropy decreases by about 5–6 times, compared with the SAR of dilute assembly of the same nanoparticles. Thus, the strong magneto-dipole interaction between the nanoparticles, generally speaking, negatively affects the ability of the assembly to absorb the energy of alternating magnetic field [1,26,27,58,59].

An important exception to this rule is a dilute oriented assembly of magnetosome chains in an alternating magnetic field applied parallel to the chain axis [44–47]. In this case the intense magnetic dipole interaction between the nanoparticles of the chain plays a positive role. Due to the influence of mutual magneto - dipole interaction within the chain, the magnetic moments of magnetically soft iron oxide nanoparticles are oriented approximately along the chain axis. In addition, the magnetization reversal of the chain in a magnetic field parallel to its axis occurs in a process similar to the giant Barkhausen jump [53]. As a result, the shape of the low-frequency hysteresis loop becomes nearly rectangular, which leads to a significant increase in the SAR of a dilute assembly of chains in comparison with the corresponding assembly of isolated magnetosomes [44]

In this paper this effect is studied in detail using numerical modeling by solving the stochastic Landau - Lifshitz equation for assemblies of magnetosome chains of various geometries. In addition to the results of previous studies [45–47], attention is paid to the correct selection of the geometric parameters of the chain, namely, the reduced distance *a/D* between the centers of the particles of the chain, consistent with the used value of the magnetic field amplitude $H_0$. It is shown that assemblies of magnetosome chains of different lengths have comparable SAR values, provided that the number of particles in the chain exceeds $N_p$ = 4–5. However, the SAR of oriented chain assembly significantly decreases for large angles $\theta$ > 50º of the magnetic field direction with respect to chain axis. In addition, the SAR of the oriented assembly of magnetosome chains also decreases rapidly with increasing average density of the assembly due to increase in the intensity of magneto – dipole interaction between the nanoparticles belonging to various chains.

## Acknowledgment

The authors gratefully acknowledge the financial support of the Ministry of Higher Education and Science of the Russian Federation in the framework of Increase Competitiveness Program of NUST «MISIS», contract № K2-2019-012.